\documentclass[aps,prl,twocolumn,showpacs,superscriptaddress]{revtex4}%
\usepackage{amsmath}
\usepackage{amsfonts}
\usepackage{amssymb}
\usepackage{graphicx}%

\begin{document}

\title{Two-gap superconductivity in single crystal Lu$_2$Fe$_3$Si$_5$ from penetration depth measurements}

\author{R. Gordon}
\affiliation{Ames Laboratory and Department of Physics \& Astronomy, Iowa State University,
Ames, IA 50011}

\author{M. D. Vannette}
\affiliation{Ames Laboratory and Department of Physics \& Astronomy, Iowa State University,
Ames, IA 50011}

\author{C. Martin}
\affiliation{Ames Laboratory and Department of Physics \& Astronomy, Iowa State University,
Ames, IA 50011}

\author{Y. Nakajima}
\affiliation{Department of Applied Physics, The University of Tokyo, Hongo, Bunkyo-ku, Tokyo 113-8656, Japan}

\author{T. Tamegai}
\affiliation{Department of Applied Physics, The University of Tokyo, Hongo, Bunkyo-ku, Tokyo 113-8656, Japan}

\author{R. Prozorov}
\email[corresponding author: ]{prozorov@ameslab.gov}
\affiliation{Ames Laboratory and Department of Physics \& Astronomy, Iowa State University,
Ames, IA 50011}

\date{31 December 2008}

\begin{abstract}
Single crystal of Lu$_2$Fe$_3$Si$_5$ was studied with the tunnel-diode resonator technique in Meissner and mixed states. Temperature dependence of the superfluid density provides strong evidence for the two-gap superconductivity with almost equal contributions from each gap of magnitudes $\Delta_1/k_BT_c=1.86$ and $\Delta_1/k_BT_c=0.54$. In the vortex state, pinning strength shows unusually strong temperature dependence and is non-monotonic with the magnetic field (peak effect). The irreversibility line is sharply defined and is quite distant from the $H_{c2}(T)$, which hints on to enhanced vortex fluctuations in this two-gap system. Altogether our finding provide strong electromagnetic - measurements support to the two-gap superconductivity in Lu$_2$Fe$_3$Si$_5$ previously suggested from specific heat measurements.
\end{abstract}


\maketitle

Originally, the interest to iron - containing silicides M$_2$Fe$_3$Si$_5$ was due to unusually high superconducting critical temperatures for compounds containing a crystallographically ordered iron sublattice \cite{Braun80}. Three silicides, all having the same tetragonal structure, have been found to be superconductors, M=Y, Sc and Lu with transition temperatures of 2.4, 4.5, and 6.0 K, respectively. It turns out that iron in these materials is nonmagnetic as was concluded from $^{57}$Fe M\"ossbauer effect measurements \cite{Cashion80,Braun81}. However, further detailed studies revealed that other superconducting properties are quite unconventional. The upper critical field H$_{c2}$(0) for  Lu$_2$Fe$_3$Si$_5$ has been found to be unusually large when compared to other iron-containing superconductors \cite{Stewart85, Umarji85} and its temperature dependence differs from conventional. The anisotropy and a pronounced peak effect in magnetic measurements was reported in \cite{Tamegai07}. The presence of a large residual electronic term in the specific heat below T$_c$ as well as a reduced specific heat jump at T$_c$ have been observed and confirmed, indicating departures from the standard BCS-like behavior \cite{Vining83,Tamegai07}. Non-magnetic impurities suppressed $T_c$ at a significant rate, incompatible with isotropic s-wave BCS picture \cite{Braun81a,Xu88}. On the other hand, ac Josephson effect indicated s-wave pairing mechanism \cite{Noer85}. Vining \emph{et al.} have proposed a two-band model in order to explain their specific heat data \cite{Vining83}. Their model assumes two-band Fermi surface with one band being superconducting and gapped, and another being normal. This represents an extreme case of multiband superconductivity as we know it today, for example in MgB$_2$ superconductor where different bands have gaps of different magnitude \cite{Bouquet01b,Manzano02}. Later detailed measurements of Lu$_2$Fe$_3$Si$_5$ crystals and analysis have shown that specific heat data are well explained quantitatively within two band model of superconductivity where both bands are gapped but with different gap amplitudes \cite{Tamegai08}.

In this letter we present precision measurements of the London and Campbell penetration depths, analyze superfluid density as well as unusual vortex properties and conclude that Lu$_2$Fe$_3$Si$_5$ is, indeed, a two-gap superconductor. It seems that multiband superconductivity is more widespread and develops when there is different dimensionality of the Fermi surface on different bands that leads to a reduced interband scattering. In MgB$_2$ there are two- and three - dimensional bands \cite{Bouquet01b,Manzano02}, whereas Lu$_2$Fe$_3$Si$_5$ has quasi-one- and three- dimensional Fermi surfaces \cite{Tamegai08}.

Measurements of Lu$_2$Fe$_3$Si$_5$ single crystal were performed using a tunnel diode resonator (TDR) \cite{Prozorov06,Prozorov00a,Prozorov00b}. Extended review of using TDR to study superconductors is given in Ref.\cite{Prozorov06}. The main components of the TDR are an LC tank circuit and a tunnel diode.  The tunnel diode has a region of negative differential resistance in its I-V curve. If a DC bias voltage is applied across the diode in this region, then it acts as an AC power source for the LC tank circuit.  This results in a self-oscillating circuit, which resonates continuously at a constant frequency for given values of L and C. The resonance frequency of the circuit used in our measurements was near 14 MHz.  All throughout the measurements the circuit is kept at a constant temperature, 4.8 K $\pm$ 1 mK, allowing for a stability of 0.05 Hz in the resonance frequency over several hours.  The sample to be studied is mounted on a sapphire rod with a small amount of Apiezon N grease. The sapphire is inserted inside of the inductor coil of the tank circuit. It is important that the sample and its mount do not make physical contact with the coil so that the temperature of the sample may be changed while keeping the circuit at a constant temperature to maintain the stability. As the magnetic susceptibility of the sample changes with temperature, so does the inductance of the tank coil. This results in a change in the TDR resonance frequency. By measuring the shift in the resonance frequency, we are able to sense changes in the penetration depth on the order of 0.5 Angstroms. Specifically, the frequency shift, $\Delta f=f\left(  T\right)  -f_{0}$, with respect to the resonant frequency of an empty coil, $f_{0}$, is given by

\begin{equation}
\Delta f\left(  T\right)  =-G4\pi\chi\left(  T\right)  =G\left[
1-\frac{\lambda}{R}\tanh\left(  \frac{R}{\lambda}\right)  \right]
\label{df}
\end{equation}

\noindent where $G\simeq f_{0}V_{s}/2V_{c}\left(  1-N\right)  $ is the geometry dependent calibration constant, $V_{s}$ is sample volume, $V_{c}$ is the
effective coil volume and $N$ is the demagnetization factor. The effective sample dimension $R$ is calculated by using Ref.\cite{Prozorov00a}. As described in detail in Ref.\cite{Prozorov00b} it is difficult to obtain the absolute value of the penetration depth due to uncertainties in the sample dimension. However, it is possible to calibrate the system with great accuracy by using temperature-dependent skin depth, $\delta\left(  T\right)  $, measured right above the $T_{c}$. In that
regime, both real and imaginary parts of the susceptibility are taken into account and the frequency shift due to skin effect is $\Delta f\left(
T\right)  _{T>T_{c}}=G\left[  1-(\delta/2R)\tanh\left(  2R/\delta\right)  \right] $. The skin depth, $\delta\left(  T\right)=c\sqrt{\rho\left(  T\right)/2\pi\omega}$ is evaluated independently from the temperature-dependent resistivity, $\rho\left(T\right)$, measured by the four-probe technique. In addition to excellent stability and sensitivity, the advantage of this technique is very low excitation fields, $\sim 20$ mOe which ensures that the sample is in Meissner state. Furthermore, by superimposing an external DC field we can probe the vortex state in so-called Campbell regime where small excitation ensures that vortices remain in their potential wells.

Single crystal of Lu$_2$Fe$_3$Si$_5$ was grown by the floating-zone technique using an image furnace followed by an annealing as described in detail elsewhere \cite{Tamegai08}. The sample was a rectangular slab having dimensions 0.99 $\times$ 0.84 $\times$ 0.15 mm$^3$ with the c-axis perpendicular to the largest face. To study possible anisotropy of the response, the measurements were performed for the excitation field both parallel and perpendicular to the c-axis of the sample. A $^3$He cryostat with sample in vacuum and external field up to 9 T was used for the reported studies.

\begin{figure}[htbp]
\begin{center}
\includegraphics{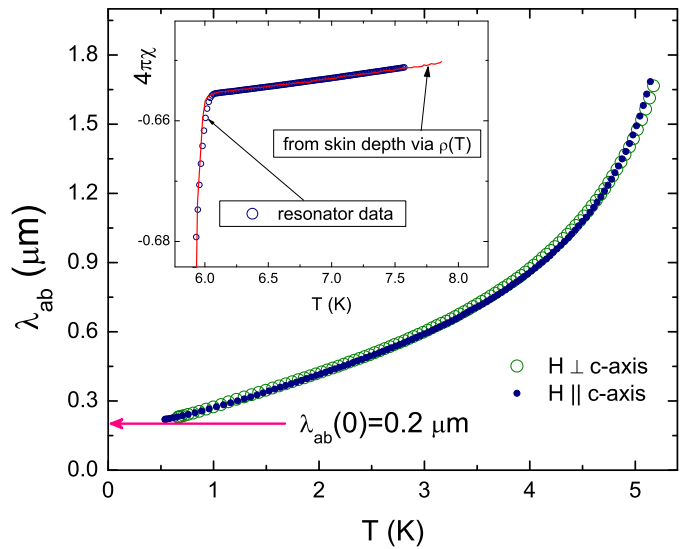}
\caption{(Color Online) $\lambda_{ab}(T)$ obtained from the measurements in two orientations. Inset: Evaluation of the calibration constant by matching $4\pi \chi (T)$ from the TDR data  (empty points) to calculated from the the skin depth (solid red curve) obtained from the resistivity.}
\label{lambda}
\end{center}
\end{figure}

Figure \ref{lambda} shows the temperature dependence of the London penetration depth, $\lambda_{ab}(T)$ obtained from the measurements along and perpendicular to the c-axis. Both orientations give $\lambda_{ab}(T)$, because sample is a thin plate and apparently $\lambda_c (T)$ is not too different from the $\lambda_{ab}(T)$ - otherwise the results would not coincide. The value of $\lambda_{ab}(0)=0.2$ $\mu$m was obtained as described in Ref.\cite{Prozorov06} from the reversible magnetization $dM/d\ln{H}$ measured independently on the same sample using \emph{Quantum Design} magnetometer. In the further analysis, possible uncertainty of this number up to 25\% was examined and and confirmed not to change our conclusions in any way.

\begin{figure}[htbp]
\begin{center}
\includegraphics{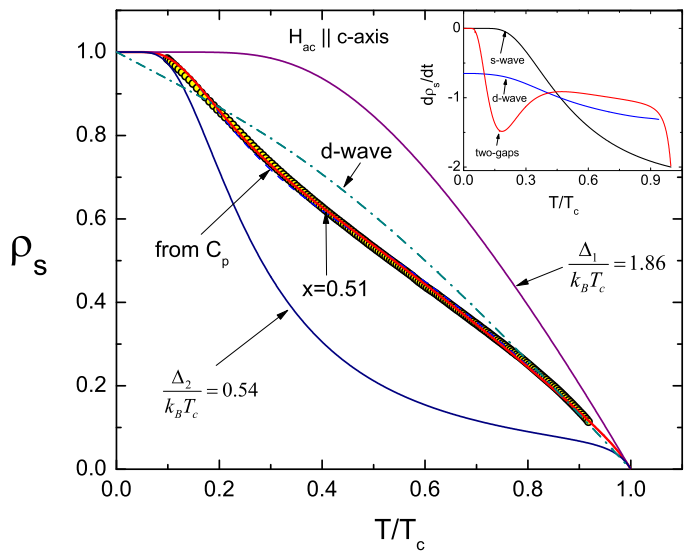}
\caption{(Color Online) TDR data (symbols) fitted to a two-gap model with indicated parameters. The red curve shows the total superfluid density. Also shown and labeled two partial superfluid densities as described in the text. The dashed line is calculated fro the parameters derived from the specific heat data. Inset: $d\rho_s/dt$ for pure d-wave, s-wave and present case of two-band superconductivity.}
\label{rho}
\end{center}
\end{figure}

Symbols in Fig.~\ref{rho} show temperature dependent superfluid density, $\rho_s(T)=\left(\lambda \left( 0 \right)/\lambda \left( T \right) \right)^2$ calculated from the penetration depth, Fig.~\ref{lambda}. The solid red curve is the total superfluid density calculated from the $\alpha$ model that assumes two independent contributions to the total superfluid density and has been successfully applied to the well known two-gap superconductor MgB$_2$ \cite{Bouquet01b,Manzano02}. In this model, each superconducting gap, $\Delta_1 (T)$ and $\Delta_2 (T)$ have similar temperature dependence given by the weak-coupling BCS self-consistency equation \cite{Prozorov06}, but with different ratio of $\Delta(0)/k_B T_c$ that become two fit parameters. A third fitting parameter gives the relative contribution of each band to the total superfluid density, $\rho_{total}(T)=x\rho_1(T)$ + (1-x)$\rho_2(T)$. Each superfluid density is calculated by using full temperature range semiclassical BCS treatment as described in detail elsewhere \cite{Prozorov06}. These partial $\rho_1(T)$ and $\rho_2(T)$ are shown by marked solid lines in Fig.~\ref{rho}. The best fit was achieved with $x=0.51$, $\Delta_1/k_BT_c=1.86$, and $\Delta_1/k_BT_c=0.54$. The first gap is quite close to the weak-coupling value of 1.76, whereas the second gap is much smaller and it is surprising that earlier two-band model assumed it to be fully normal \cite{Vining83}. Similarly to MgB$_2$, the two gaps contribute equally to the superfluid density. A dashed line, which almost follows the data is calculated from the parameters obtained analyzing specific heat data, $x=0.47$, $\Delta_1/k_BT_c=2.2$, and $\Delta_1/k_BT_c=0.55$ \cite{Tamegai08}, which is in a quite good agreement given very different nature of the measurements. To further highlight the qualitative difference between single and two gap behavior, we plot $d\rho_s/dt$ in the inset to Fig.~\ref{rho}. Note characteristic nonmonotonic behavior in the case of two gaps. It is not present either in pure d-wane, nor pure s-wave cases.

\begin{figure}[htbp]
\begin{center}
\includegraphics{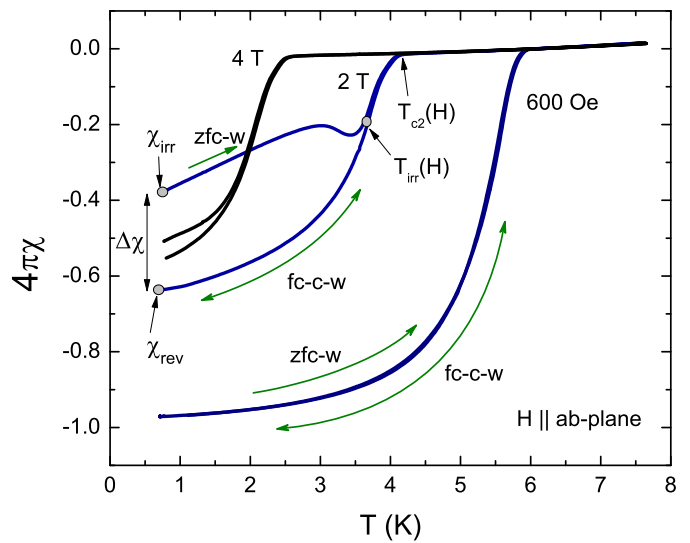}
\caption{(Color Online) $4\pi \chi(T)$ from the TDR data at three different values of an applied magnetic field along the c-axis. Each curve was obtained after cooling in zero field and then warming and cooling twice. The labels and arrows indicate various characteristic points used in later analysis.}
\label{zfcfc}
\end{center}
\end{figure}

Whilst the situation is quite clear for the London penetration depth, measurements in an applied magnetic field reveal more puzzling behavior of the studied compound. When an external DC field is applied and small - amplitude AC response is probed, vortices respond elastically and the overall susceptibility is governed by the Campbell penetration depth, $\lambda^{2}=\lambda_{L}^{2}+\lambda_{C}^{2}$, where $\lambda_{L}$ is the usual London penetration depth described above and $\lambda_{C}(B,T,j)$ is the Campbell penetration depth \cite{Prozorov03}, $\lambda_{C}^{2}=\phi_{0}B/{4\pi\alpha(j)}$. Here $\phi_{0}$ is the flux quanta and ${\alpha(j)}$ is the Labusch parameter that generally depends on the biasing Bean current generated in the sample, for example, after applying field after cooling in zero field. Sample magnetic susceptibility (and the frequency shift) in the vortex state is still given by Eq.~\ref{df}, but with generalized penetration depth.

In conventional type-II superconductors there is no hysteresis for zero-field cooled (zfc) and field cooled (fc) curves of the small amplitude AC response. However, in materials where $j_c$ is strongly temperature dependent (e.g. high-$T_c$ cuprates) large hysteresis is observed \cite{Prozorov03}. As shown in Ref.\cite{Prozorov03}, cubic correction to a parabolic potential well for vortex pinning leads to ${\alpha(j)=}\alpha_{0}\sqrt{1-j/j_{c}}$, where $j_{c}=c\alpha_{0}r_{p}/{\phi_{0}}$ is the critical current and $r_{p}$ is the radius of the pinning potential. This model explains why zero-field cooled curve differs from subsequent cooling and warming and it was successfully used to explain the data for Bi$_2$Sr$_2$CaCu$_2$O$_{8+y}$ superconductor.

$4\pi \chi(T)$ in the vortex state of Lu$_2$Fe$_3$Si$_5$ is shown in Fig.~\ref{zfcfc} for three representative fields. In each case, sample was cooled in zero applied field to the base temperature and indicated magnetic field was applied. Then measurements were taken while warming up the sample above $T_c$ (zfc-w). Then, sample was cooled and warmed twice without changing the field and while taking the data (fc-c and fc-w). For low field values, there is no hysteresis observed, while at intermediate fields the hysteresis becomes very pronounced. Clearly, the hysteresis is associated with the static Bean current, $j$, induced by applying field. We also note that this effect is not associated with the vortex density (e.g. less vortices after zfc), because then initial Campbell length would be smaller than equilibrium, not larger as observed.

\begin{figure}[htbp]
\begin{center}
\includegraphics{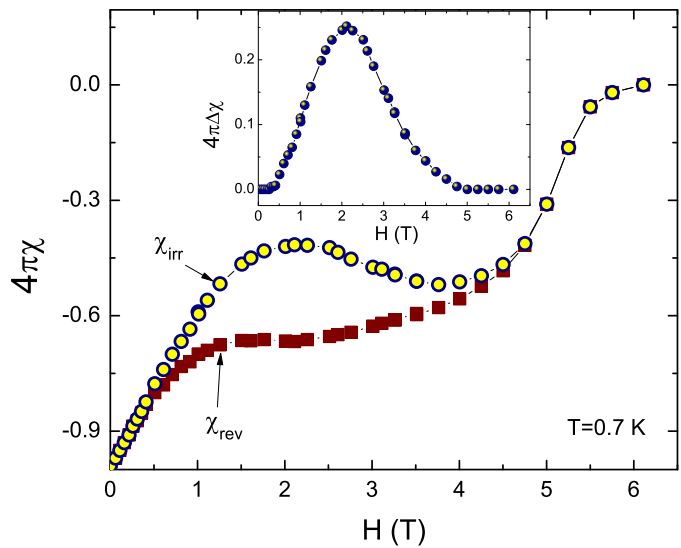}
\caption{(Color Online) Open circles: $4\pi \chi_{irr}(H)$ at $T=0.7$ K measured by applying field after zfc as indicated in Fig.~\ref{zfcfc}. Closed squares indicate $4\pi \chi_{rev}(H)$ obtained on fc. Inset: the difference between zfc and fc curves shown in the main frame, $4 \pi \Delta \chi =4 \pi \chi_{irr}-4 \pi \chi_{rev}$.}
\label{fishtail}
\end{center}
\end{figure}

By measuring many $4\pi \chi(T)$ curves at different magnetic fields, we extracted field dependence of the initial susceptibility obtained after zfc and fc. Figure \ref{fishtail} shows the resulting $4\pi \chi_{irr}(H)$ (open circles) and $4\pi \chi_{rev}(H)$ (closed squares) curves at $T=0.7$ K. The inset shows the difference between the two curves. This difference is directly related to the strength of pinning and magnitude of the apparent Bean current density, $j$, $\Delta \chi \sim j/j_c$ where we assumed $j \ll j_c$. There is a clear peak effect and its location is well compatible with direct measurements reported in Ref.~\cite{Tamegai07}.

\begin{figure}[htbp]
\begin{center}
\includegraphics{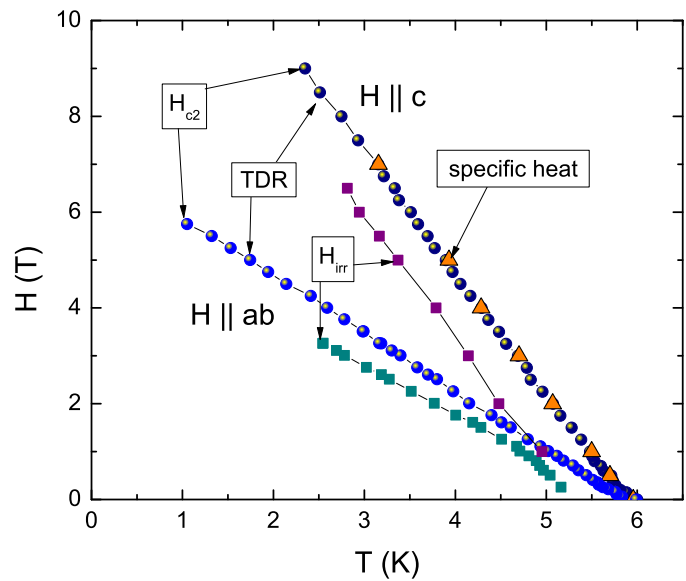}
\caption{(Color Online) $H-T$ phase diagram for Lu$_2$Fe$_3$Si$_5$ crystal in two orientations.}
\label{ht}
\end{center}
\end{figure}

Finally, we construct the $H-T$ phase diagram obtained from our measurements for both directions. While Meissner response is governed by currents flowing in the ab-plane, in magnetic field the response is anisotropic and is determined by orientation of vortices with respect to crystal axes. We observe large anisotropy of the upper critical field, $H_{c2}(T)$, down to 1 K as shown in Fig.~\ref{ht}, which has not been reported in earlier papers. Furthermore, $H_{c2}(T)$ determined from the TDR measurements is in excellent agreement with the specific heat data. Note that $H_{c2}(T)$ is linear in temperature  down to $0.15T_c$. Figure \ref{ht} also shows position of the irreversibility line (see Fig.~\ref{zfcfc} for definition) for both orientations. Unlike conventional superconductors where $H_{irr}(T)$ is very difficult to determine, because it gradually merges into $H_{c2}(T)$, in Lu$_2$Fe$_3$Si$_5$ it is sharply defined and is quite distant from the $H_{c2}(T)$, which is another indication of significant reduction of the critical current possibly due to enhanced fluctuations in the two-gap system.

In conclusion, we find that Lu$_2$Fe$_3$Si$_5$ shows Meissner response compatible with two-gap s-wave superconductivity, similar to MgB$_2$. It seems that having Fermi surfaces of different dimensionality is the important ingredient for multi-gap superconductivity. In the vortex state, Lu$_2$Fe$_3$Si$_5$ shows unusually strong temperature dependence of the critical current, which is also non-monotonic with magnetic field (peak effect). The upper critical field is anisotropic and linear in temperature. All these observations are reminiscent of unconventional superconductivity and further theoretical insight to connect these properties is needed.

Discussions with P. C. Canfield and V. G. Kogan are greatly appreciated. Work at the Ames Laboratory was supported by the Department of Energy-Basic Energy Sciences under Contract No. DE-AC02-07CH11358. Work at the University of Tokyo was supported by a Grant-in-Aid for Scientific Research from the Ministry of Education, Culture, Sports, Science and Technology. R. P. acknowledges partial support from NSF grant number DMR-05-53285 and the Alfred P. Sloan Foundation.

\end{document}